\documentclass[conference]{IEEEtran}

\usepackage[letterpaper,margin=1in,includefoot]{geometry}
\IEEEoverridecommandlockouts
\usepackage{cite}
\usepackage{amsmath,amssymb,amsfonts}
\usepackage{graphicx}
\usepackage{textcomp}
\usepackage{subcaption}
\usepackage[table,xcdraw]{xcolor}
\usepackage[]{algorithm2e}
\def\BibTeX{{\rm B\kern-.05em{\sc i\kern-.025em b}\kern-.08em
    T\kern-.1667em\lower.7ex\hbox{E}\kern-.125emX}}
\usepackage{multirow}
\usepackage{colortbl}
\usepackage{hhline}
\usepackage{tikz}
\usepackage{amssymb}
\usepackage{pifont}
\usepackage{diagbox}
\usepackage{url}
\usepackage{verbatim}
\usepackage{soul}

\usepackage{booktabs} 
\usepackage[symbol]{footmisc}
\usepackage{url}

\usepackage{hyperref,xcolor}

\hypersetup{
  colorlinks
}

\newcommand{\percent}[1]{{$#1\%$}}

\newcommand{\nvj}{NVJPEG}
\newcommand{\nvjL}{NVJPEG hardware unit}

\newcommand{\orinL}{Jetson AGX Orin}

\newcommand{\pypre}{PyPre}
\newcommand{\pypreL}{PyTorch-Preprocessing}

\newcommand{\cdpre}{CDPre}
\newcommand{\cdpreL}{CPU-DALI-Preprocessing}

\newcommand{\cdpredla}{CDPre-DLA}
\newcommand{\cdpredlaL}{CPU-DALI-Preprocessing with DLA inference}

\newcommand{\ndpre}{NDPre}
\newcommand{\ndpreL}{NVJPEG-DALI-Preprocessing}

\newcommand{\mi}{MI}
\newcommand{\miL}{Multi-Instance Design without NVJPEG}

\newcommand{\minv}{MI-NVJ}
\newcommand{\minvL}{Multi-Instance Design with NVJPEG}

\usepackage{makecell}

\usepackage{amsmath}

\usepackage{pifont}

\usepackage{url}

\begin{document}

\title{Accelerating Data Preprocessing for Efficient \\ Vision Model Inference on Jetson Edge Device}

\author{
	Tian Chen, Nawras Alnaasan, Jinghan Yao, Aamir Shafi, Hari Subramoni, and \\ Dhabaleswar K. (DK) Panda \\
	The Ohio State University\\
    
    \begin{normalsize}
        \begin{sffamily}
            \{chen.9891, alnaasan.1, yao.877, shafi.16, subramoni.1, panda.2\}@osu.edu
        \end{sffamily}
    \end{normalsize}

}

\maketitle
\pagestyle{plain}
\begin{abstract}
Data preprocessing is a crucial part of deep learning workflows on edge devices. However, decoding data saved in JPEG format is very compute-intensive and occupies a major portion of the preprocessing pipeline. Therefore, increasing the decoding speed is vital for improving overall throughput, especially for inputs with large image sizes, which are often subject to preprocessing bottlenecks. 
On the other hand, edge devices are equipped with specialized hardware units to accelerate media processing and image decoding. For instance, the NVIDIA Jetson platform possesses a dedicated NVJPEG unit. These units can be used to enhance the performance of the preprocessing pipeline. 

This paper introduces the utilization of such specific hardware acceleration units for offloading decoding tasks. By combining this with a multi-instance approach, it allows for the parallelization of all compute resources including CPU, NVJPEG, GPU, and DLA in Jetson devices. In this work, we compare various potential pipeline designs.
On ResNet18, ResNet50, and ResNet152, three models with different sizes, we evaluate the impact of batch sizes and image sizes, as well as the characteristics of GPU/DLA inference. 
Finally, a fine-tuning experiment for multi-instance design has been conducted. The multi-instance design with a specific hardware decoding unit involved offers up to \percent{30.02} speedup for large image sizes, compared with the most optimized design without it. Based on these findings, we demonstrate the benefits of using the NVJPEG unit in deep learning workflows and provide guidelines for tuning and optimizing edge inference workflows.

\end{abstract}


\vspace{-4.0ex}

\section{Introduction}
\label{sec:intro}

Nowadays, Deep Learning (DL) has achieved remarkable success across a wide range of domains. Although deploying DL models in cloud can provide users with convenient and accessible experiences, transmitting data from edge devices to the cloud for inference introduces several challenges, including network latency, resource scalability, and data privacy concerns~\cite{chen_deep_2019}. Due to the need for real-time processing and/or privacy considerations, the deployment of DL models on edge devices for inference is becoming increasingly popular. Common applications of edge DL include autonomous driving, natural language processing, computer vision, and more.

Vision DL models are a prominent type of application in edge inference, encompassing tasks such as object detection, classification, image segmentation, and semantic recognition. These tasks require edge devices to process data from cameras in real-time,  undergoing various transformations before the images are suitable for consumption by the DL model. For edge devices with limited computational resources, such data preprocessing is a vital step. This step, along with inference speed, is crucial in determining the overall throughput.

\subsection{Problem Statement}

Data preprocessing for vision DL models is a crucial yet resource-intensive step. To save storage space and transmission bandwidth, images are often saved and provided in a compressed format. JPEG, for instance, is a widely used compressed image format renowned for its efficiency in saving space and enabling fast transmission. However, decoding JPEG images can consume a significant portion of the preprocessing time in workflows.

As demonstrated in~\ref{sec:ComparingPreprocessingPipelines} Fig.~\ref{fig:BSChart_lat}, when using the state-of-the-art PyTorch~\cite{pytorch} framework with ResNet50~\cite{resnet} for inference over the ILSVRC 2012~\cite{imagenet} dataset, preprocessing can account for up to \percent{64} of the entire workflow. A further breakdown of preprocessing time in~\ref{sec:breakdownEval} reveals that image decoding can take from \percent{52.54} to \percent{87.30} of the total preprocessing time, depending on the image size. It shows image decoding already becomes a bottleneck that slows down the entire pipeline's throughput, especially with large image sizes. Therefore, efficiently enhancing the JPEG decoding process becomes crucial for optimizing edge inference throughput.

\subsection{Motivation}

There is a pressing need to address the bottleneck in image decoding during data preprocessing, leveraging specialized hardware units in edge devices, which are designed for efficient media and image processing, is a viable solution. Offloading image decoding to these specialized units can substantially improve efficiency. However, this approach alone is insufficient for maximizing the use of all compute resources in the device to enhance inference throughput. Since when these specialized units process data, the CPU's computational resources often remain underutilized. Therefore, it's imperative to develop a methodology that can concurrently engage all available units within the edge device for optimal performance.

Parallel and pipeline are two common approaches in optimizing edge DL workflows. By inserting intermediate buffers between pipeline stages, implementing multi-threaded pre/post-processing, and employing model parallelism across GPUs and DLAs~\cite{jeong_tensorrt-based_2022}, hardware utilization can be enhanced. However, there is a lack of existing work that considers the bottleneck in preprocessing and the benefits of specialized hardware units in accelerating decoding. Moreover, considering the unique characteristics of edge devices, there is no one-size-fits-all solution for different model sizes and datasets with various image sizes. Therefore, there is a necessity for a systematic study in this effort. In this paper, we aim to efficiently leverage all hardware units in edge devices simultaneously to accelerate both data preprocessing and inference, thereby speeding up the entire edge DL workflow.

\subsection{Contributions}

In this study, we propose the integration of NVJPEG, a hardware-accelerated JPEG decoding unit in~\orinL{}~(Orin), into the data preprocessing pipeline of vision DL workflow. This integration, coupled with a multi-instance design, enables the effective utilization of all hardware components of the device. Additionally, we have conducted a comprehensive evaluation to understand the characteristics of the Orin device. This study also provides a set of tuning guidelines to optimize performance.

The paper makes the following key contributions:

\begin{itemize}
    
    \item We propose using a novel NVJPEG with a multi pipeline instances design to utilize all hardware components including NVJPEG, CPU, GPU, and Deep Learning Accelerator(DLA). Compared with the most optimized design without NVJPEG involved, it delivers up to \percent{30.02} speed up for large image size scenarios. 
    
    \item We evaluate the characteristics of the preprocessing pipeline varying batch size and image size. A time breakdown evaluation has been conducted to explore the time constitution of the preprocessing pipeline. 

    \item Study the effectiveness of multi-instance design is delivered, showing the effectiveness of increasing pipeline instances. We evaluate the characteristics of each hardware unit of the Orin device, including a comparison of the CPU and NVJPEG for decoding and the GPU and DLA for inference.

\end{itemize}

\section{Background}
\label{sec:bgnd}

\subsection{NVIDIA Jetson AGX Orin}

In this work, we conducted our work on \orinL{}. It is a representative edge device that has been widely deployed for various tasks. It contains a 12-core Arm Cortex-A78AE at 2.2GHz CPU, one 1.3GHz NVIDIA Ampere architecture-based GPU with 2048 CUDA cores and 64 Tensor Cores. To leverage DL workflow, it contains two NVIDIA Deep Learning Accelerators v2.0 at 1.6GHz, together deliver a peak of 105 TOPs INT8 performance. To handle media processing, it is equipped with a Programmable Video Accelerator, NVIDIA Video Encoder/Decoder, and NVJPEG hardware unit. All those units are connected via a memory controller fabric to 32GB LPDDR5 unified memory.

\subsection{Data Preprocessing in Vision Models}

Data preprocessing, especially for Vision models, will involve compute-intensive image processing and transformation. Raised a challenge for edge DL workflow. 
A notable approach is offloading data transformation and augmentation to GPU like what NVIDIA's Data Loading Library~\cite{nvidia_dali} (DALI) will do. It significantly cuts the latency associated with CPU processing.

Besides transformation, JPEG decompression will contribute to a large portion of time in the data preprocessing step. During JPEG decompression, data is divided into many Minimum Coded Units (MCU). A larger image will have more MCU, leading to more computing to decode it. The compression of JPEG happens in YUV color space, based on the fact that the human eye is less sensitive to color than lightness. Meanwhile, the DL model usually takes images in RGB color space as input instead of YUV space. Therefore, a color space conversion is required before feeding the decompressed JPEG into the DL model.

\subsection{Edge Inference}

Vision DL models are widely adopted in edge devices. Inference in edge device leverages technologies like NVIDIA TensorRT~\cite{tensorRT}. TensorRT optimizes DL models for high-performance inference, reducing latency and improving throughput. A typical TensorRT workflow will include taking the DL model from the training framework, building an optimized TensorRT engine, and loading the engine for execute inference. For GPU and DLA, TensorRT will build different engines optimized for them separately. Quantization is a common approach during engine building. Computations can be performed faster by using lower precision datatypes such as INT8 for quantization in hardware.

\section{design}

In this section, we initially introduce various pipeline designs, followed by a detailed analysis of their effectiveness. Finally, we employ a timeline breakdown approach to explore the drawbacks of existing methods and the benefits of our proposed approaches.

\begin{table}[ht]
\centering
    \vspace{-1.5ex}
    \renewcommand{\tabcolsep}{3pt}
\caption{Evaluated Pipeline Designs}

\label{tab:designs}
\begin{tabular}{@{}llll@{}}
\toprule
  \textbf{Pipeline Design}
&  \makecell{\textbf{Preprocessing} \\ \textbf{framework} }
&  \makecell{\textbf{Decoding} \\ \textbf{Device} }
&  \makecell{\textbf{Inference} \\ \textbf{Device} }
\\
\midrule
\midrule

\pypre  & PyTorch & CPU & GPU \\ \hline
\cdpre & DALI    & CPU & GPU \\ \hline

\cdpredla  & DALI    & CPU & DLA \\ \hline

\mi  & DALI    & CPU & GPU + DLA \\ \hline

\textcolor{red}{\ndpre{} (Proposed)}  & DALI    & NVJPEG & GPU \\ \hline
\textcolor{red}{\minv{} (Proposed)}  & DALI    & CPU + NVJPEG & GPU + DLA \\ \hline

\bottomrule
\end{tabular}
  \vspace{-2ex}
\end{table}

\subsection{Evaluated Pipeline Designs}

This section outlines six pipeline designs utilized in the experiment, as detailed in Table~\ref{tab:designs}. Each DL workflow is divided into preprocessing and inference pipelines, with image decoding on CPU/NVJPEG and inference on GPU/DLA—GPU by default unless named with a `DLA' suffix:
\begin{enumerate}

\item{
\textbf{\pypreL{} (\pypre{}):}
Using PyTorch with the CPU for preprocessing. 
}

\item{
\textbf{\cdpreL{} (\cdpre{}):}
Using DALI with the CPU for decoding, GPU for transformation. 
}

\item{
\textbf{\cdpredlaL{} (\cdpredla{}):}
    Same setup as \cdpre{} but use DLA for inference. Notice only one DLA will be used per pipeline instance. 
}

\item{
\textbf{\miL{} (\mi{}):}
Consist with pipeline instances of \cdpre{}, \cdpredla{}. Batch size for DLA and GPU pipeline is optimized seperately. 
}

\item{
\textbf{\ndpreL{} (\ndpre{}):}
Similar to \cdpre{}, but using \nvj{} to offload JPEG decoding.
}

\item{
\textbf{\minvL{} (\minv{}):}
Similar to \mi{}, but forcing to include one NDPre pipeline instance. Could benefit from potential overlapping of CPU and NVJPEG decoding. 
}

\end{enumerate}

\begin{figure}[ht]
     \centering
     
\begin{subfigure}[b]{\columnwidth}
    \centering
    \includegraphics[width=\linewidth]{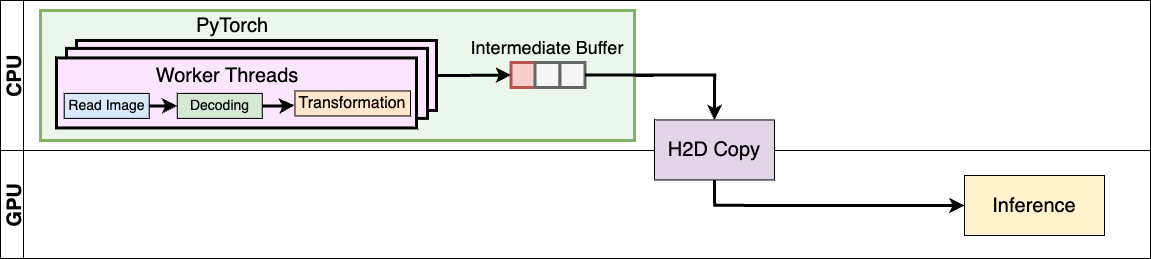}
    \vspace{-1.5em}
    \caption{\pypreL~(\pypre) workflow}
    \label{fig:design:PyPre}
\end{subfigure}

\begin{subfigure}[b]{\columnwidth}
    \centering
    \includegraphics[width=\linewidth]{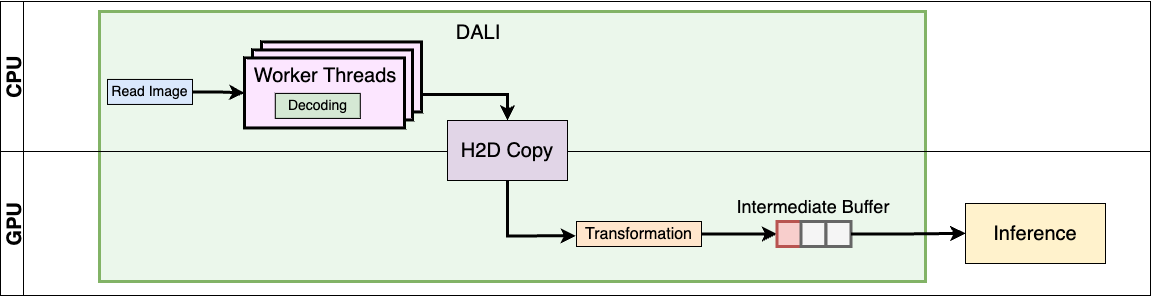}
    \vspace{-1.5em}
    \caption{\cdpreL~(\cdpre) workflow}
    \label{fig:design:CD}
\end{subfigure}

\begin{subfigure}[b]{\columnwidth}
    \centering
    \includegraphics[width=\linewidth]{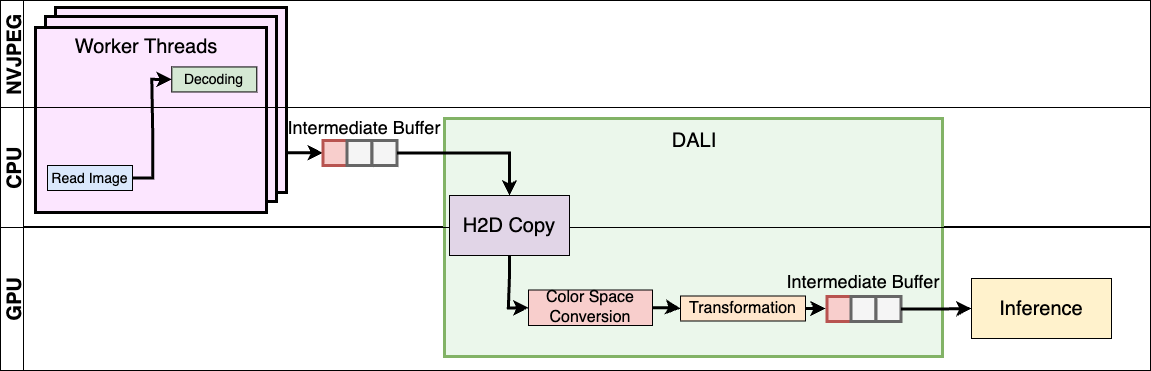}
    \vspace{-1.5em}
    \caption{\ndpreL~(\ndpre) workflow}
    \label{fig:design:ND}
\end{subfigure}

\vspace{-1.2ex}
\caption{Workflow of different single-instance preprocessing pipelines. Each lane indicates a hardware unit, with the timeline progressing from left to right.}
\vspace{-0.6em}

\label{fig:design:pipeline}
\end{figure}
 
\label{sec:design:prepipeline}

\subsection{Preprocessing Pipelines}

In a deep learning inference workflow, a preprocessing pipeline is required to take a batch of image files, such as JPEG, as input and generate tensors with model-specific transformations to feed into the inference pipeline. Such a process typically includes three main components: reading the image, decoding, and transformation. In this paper, we compared three potential preprocessing pipelines. 

\subsubsection{PyTorch-Preprocessing (PyPre)}
PyTorch is a widely used state-of-the-art deep learning framework that offers handy dataloader and Torchvision tools for the preprocessing process. As shown in Fig.\ref{fig:design:PyPre}, all three components in PyTorch are executed by the CPU. PyTorch employs multiple worker threads for parallel processing to enhance performance and features an intermediate buffer for prefetching. This setup allows the CPU to preprocess the next batch when the GPU is performing inference, thereby increasing efficiency. 

\subsubsection{CPU-DALI-Preprocessing (CDPre)}
For image processing, GPUs outperform CPUs in efficiency. NVIDIA DALI leverages GPU acceleration for efficient image transformation pipelines, as depicted in Fig.\ref{fig:design:CD}. Another benefit is that the output of the DALI pipeline is a GPU tensor that can be directly used for inference, eliminating the need for a host-to-device memory copy.

\subsubsection{NVJPEG-DALI-Preprocessing (NDPre)}

In both PyPre and CDPre, JPEG decoding takes considerable CPU time. Therefore, we propose using the NVJPEG hardware unit to offload these operations. This approach involves a more fine-grained pipeline that runs NVJPEG operations in parallel with other devices, as shown in Fig.\ref{fig:design:ND}. Our proposed NVJPEG-DALI-Preprocessing pipeline utilizes multiple worker threads to increase the utilization rate of the NVJPEG unit. It employs an intermediate buffer to decouple it from the DALI pipeline.

\def\TimeLineSpaceBEFORE{-4.5ex}
\def\TimeLineSpaceBEFOREii{-5.5ex}
\def\TimeLineSpaceAFTER{-2ex}

\begin{figure*}[t]
     \centering
     
\begin{subfigure}[b]{\textwidth}
    \centering
    \includegraphics[width=\linewidth]{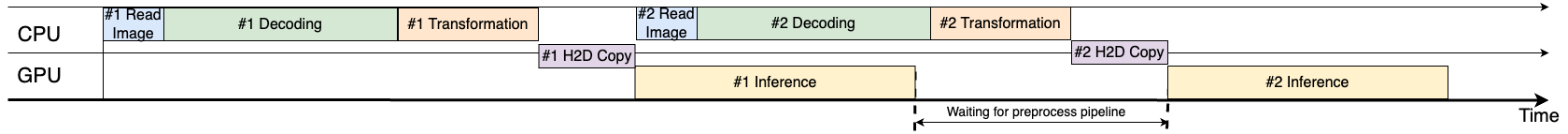}
    \vspace{\TimeLineSpaceBEFORE}
    \caption{\pypreL}
    \label{fig:design:PyPre_timeline}
\end{subfigure}
\vspace{\TimeLineSpaceAFTER}

\begin{subfigure}[b]{\textwidth}
    \centering
    \includegraphics[width=\linewidth]{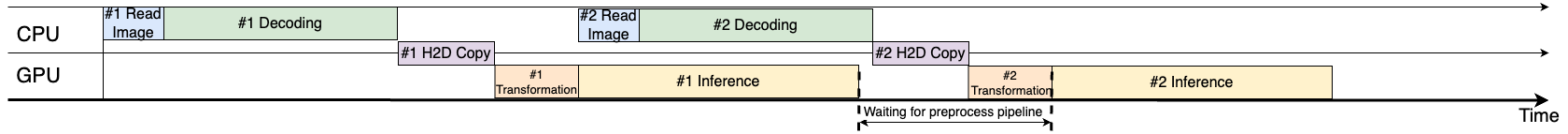}
    \vspace{\TimeLineSpaceBEFORE}
    \caption{\cdpreL}
    \label{fig:design:CDPre_timeline}
\end{subfigure}
\vspace{\TimeLineSpaceAFTER}

\begin{subfigure}[b]{\textwidth}
    \centering
    \includegraphics[width=\linewidth]{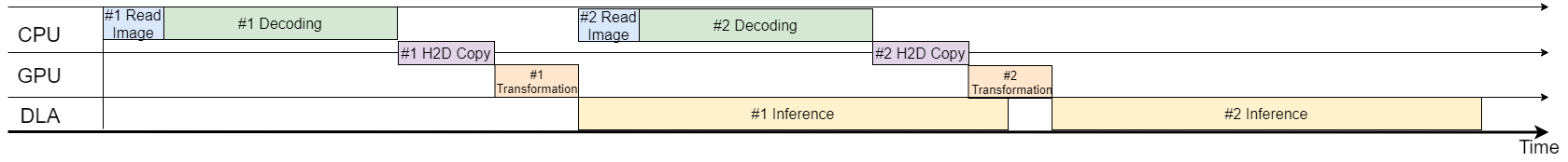}
    \vspace{\TimeLineSpaceBEFORE}
    \caption{\cdpredlaL}
    \label{fig:design:CDPredla_timeline}
\end{subfigure}
\vspace{\TimeLineSpaceAFTER}

\begin{subfigure}[b]{\textwidth}
    \centering
    \includegraphics[width=\linewidth]{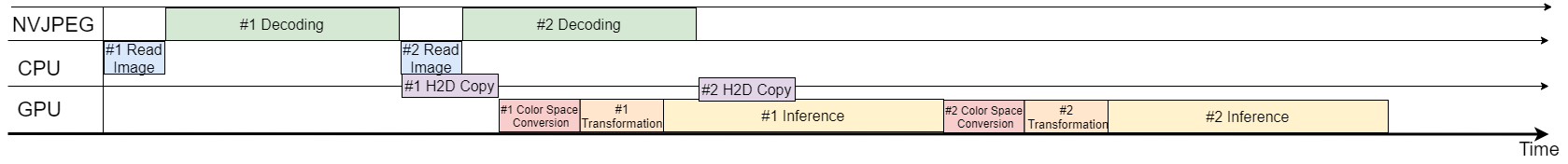}
    \vspace{\TimeLineSpaceBEFOREii}
    \caption{\textcolor{red}{\ndpreL{} (Proposed)}}
    \label{fig:design:NDPre_timeline}
\end{subfigure}
\vspace{\TimeLineSpaceAFTER}

\begin{subfigure}[b]{\textwidth}
    \centering
    \includegraphics[width=\linewidth]{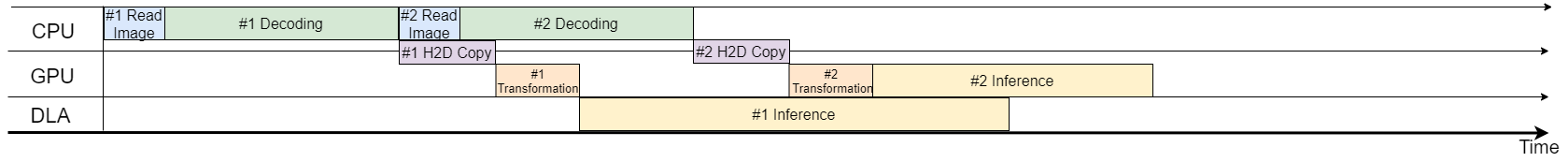}
    \vspace{\TimeLineSpaceBEFOREii}
    \caption{\miL}
    \label{fig:design:MI_timeline}
\end{subfigure}
\vspace{\TimeLineSpaceAFTER}

\begin{subfigure}[b]{\textwidth}
    \centering
    \includegraphics[width=\linewidth]{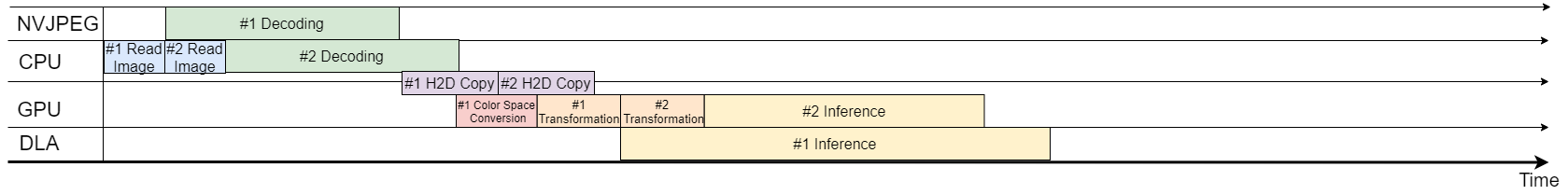}
    \vspace{\TimeLineSpaceBEFOREii}
    \caption{\textcolor{red}{\minvL{} (Proposed)}}
    \label{fig:design:MI-NVJ_timeline}
\end{subfigure}
\vspace{\TimeLineSpaceAFTER}

    \caption{Timeline for different designs. Each lane represents a hardware unit. \#1 and \#2 indicate two different batches}
    \label{fig:design:pipeline_timeline}
\end{figure*}

\subsection{Multi-Instance}

Given that the Orin device has multiple hardware computing units, including CPU, GPU, DLA, and NVJPEG, the multi-instance design aims to utilize all these units concurrently during the deep learning workflow to enhance overall throughput. 

For single-instance designs, there is a limitation that they cannot simultaneously utilize CPU and NVJPEG for decoding, or GPU and DLA together for inference. In contrast, multi-Instance approach allows for utilizing all available hardware devices. In addition, it will enable the allocation of the most optimal batch size for both DLA and GPU separately.

Furthermore, compared to single-instance scenarios, duplicating instances can enhance device utilization. This approach allows the device to execute another instance while waiting for the completion of one instance on other devices. For example, while the GPU waits for the transformation of a second instance, it can concurrently perform inference on the first instance. 

For the number and type of instances in MI design, careful selection is crucial. In scenarios where preprocessing is the limiting factor, particularly with large input image sizes, combinations of NVJPEG and CPU instances are preferable. However, the total number of instances should be carefully considered. Excessively creating instances can introduce additional overhead, negatively affecting performance. Moreover, as the number of instances increases, each must reduce its batch size to avoid memory exhaustion, which can detrimentally affect processing throughput.

On the contrary, in inference-bounded scenarios, it's important to fully utilize the computational power of the two DLAs and one GPU available in the Orin device. A combination of all available inference devices should be chosen. Additionally, the batch size for the DLA should be carefully selected, as the DLA has a limited hardware buffer. Too large a batch size could lead to a degradation in DLA performance.

\subsection{Timeline Analysis}

\subsubsection{Timeline Breakdown Modeling}
\label{sec:design:timeline:formula}

The preprocessing pipeline time consists of three parts: reading images, decoding, and transformation. For both \cdpre{} and \ndpre{} pipelines, they share the same GPU transformation portion, which consists of image resize, central clip, normalization, and datatype casting.

For \cdpre{}, the entire time used can be formulated as follows: 

\vspace{-4ex}
\begin{align}
T_{CDPre} &= BS* (T_{Read} + \frac{T_{CPU\ Decode}}{N_{threads}} + T_{H2D})\notag\\
&  + T_{Transformation}
\label{eq:cdlat}
\end{align}
\vspace{-4ex}

Where the $BS$ stands for Batch Size. Each image in batch requires a JPEG image read operation with time $T_{Read}$, a JPEG to RGB array decoding time $T_{CPU\ Decode}$, followed by a Host to Device memory copy $ T_{H2D}$. CPU decoding happens in a multi-thread manner with thread number of $N_{threads}$, each thread handles one image at the same time. 

For \ndpre{}, the formula has minor change compare with \cdpre{} which can be written as:

\vspace{-3ex}
\begin{align}
T_{NDPre} &= BS * T_{Read} + BS * \frac{T_{NVJPEG\ Decode}}{T} \notag \\
& + BS * T_{H2D} + T_{Transpose} \notag \\
& + T_{Color\ Space} * BS + T_{Transformation}
\label{eq:ndlat}
\end{align}
\vspace{-4ex}

Where the $T_{Read}$, $ T_{H2D}$ is identical to \cdpre{} formula, while decoding Tn happens in \nvjL{}. The output of \nvjL{} is in YUV color space with 4:2:0 format, which requires two additional GPU kernels: a batch transpose with time $T_{Transpose}$ and color space conversion with time $T_{Color\ Space}$ for each image.

\subsubsection{Timeline Comparison over Different Designs}

Fig.~\ref{fig:design:PyPre_timeline} shows that \pypre{}, used as the baseline design, includes an intermediate buffer. This buffer enables the GPU to perform inference on the current batch while the CPU fetches the next batch, allowing for overlapping operations. The primary objective in optimizing the preprocessing pipeline is to minimize the time waiting for the preprocessing pipeline, which refers to the duration the inference device is idle, awaiting the next input batch ready. 

Notice that in the \pypre{} design, each inference batch incurs a preliminary Host-to-Device (H2D) Copy overhead. In contrast, for \cdpre{}, the transformation happens in the GPU, which is much faster than the CPU. The H2D copy would happen before transformation and overlaps with GPU inference.

For \cdpredla{}, the inference is offloaded to the DLA. This offloading ensures that the inference runs without disrupting the transformation of the next batch on the GPU. Such a setup allows for the potential overlap between inference and transformation, proving especially advantageous in scenarios requiring extensive transformations.

Fig.~\ref{fig:design:NDPre_timeline} illustrates the timeline of \ndpre{} design. An intermediate buffer is used to achieve a more fine-grained pipeline as proposed. This design decouples NVJPEG decoding from the DALI pipeline, allowing the NVJPEG hardware unit to offload decoding while overlapping with all other operations. While the extra overhead caused by color space conversion might make this approach less efficient than \cdpre{} in GPU-bounded scenarios or involving small image sizes where decoding does not constitute a significant portion of the processing time.

Fig.~\ref{fig:design:MI_timeline} presents the multi-instance design. Compared to the single-instance designs discussed earlier, multi-instance enables different instances to operate interleaved. This effect is akin to having a more fine-grained pipeline. However, compared with the fine-grained pipeline, multi-instance design can include more inference units, fully leveraging the power of GPU and DLAs available on the Orin device. Boost compute-bounded scenarios.

Notice that in the \ndpre{} design, the CPU is underutilized, idling while waiting for the NVJPEG to return results. This dropback could be exactly addressed by combining with a multi-instance approach. As shows in Fig.~\ref{fig:design:MI-NVJ_timeline}, as enhancement to \mi{} design, \minv{} not only DLA and GPU, but CPU and NVJPEG can all overlap with each other. It boosts performance in preprocessing bounded situations, especially for large input image size.

\section{Evaluation and analysis}
\label{sec:evaluation}

\subsection{Experimental Setup}
\label{sec:setup}

\textbf{Hardware and Software:}
All experiments are conducted on a \orinL{} device with JetPack SDK~5.1.1~\cite{noauthor_jetpack_nodate} installed. JetPack SDK packed with TensortRT~8.5.2.2~\cite{tensorRT} inference framework and CUDA~11.7\cite{cuda}. 
Orin device is set to MAXN power mode, \textit{jetson\_clocks} command is used to pin the device into maximum frequency. 

Support for DALI~\cite{nvidia_dali} on Jetson devices is currently in an experimental stage. DALI 1.29.0.dev is compiled from the source code using the officially recommended build flags for the Jetson platform. At the time breakdown profiling experiment, DALI will be recompiled with NVTX~\cite{noauthor_nvidianvtx_2023} support. An NVIDIA build of PyTorch 2.0.0~\cite{noauthor_installing_nodate} is installed. 

To utilize the NVJPEG hardware unit, a customized C-written Python library is used, which wraps Jetson Linux Multimedia API (MMAPI)~\cite{noauthor_jetson_nodate}. MMAPI is a collection of lower-level APIs that provide control over underlying hardware, including NVJPEG, NVDEC Video Decoder, Programmable Vision Accelerator, and others. Within this library, we implemented a producer-consumer queue with an intermediate buffer. This design allows the NVJPEG hardware unit to decode in an asynchronous manner, enhancing the potential of overlapping tasks across different devices.

\textbf{Dataset:}
The ImageNet Large Scale Visual Recognition Challenge 2012 (ILSVRC 2012)~\cite{imagenet} is a widely used dataset that supports various tasks such as object classification, detection, and localization, aligning well with the nature of tasks for edge devices. In the experiment, the ILSVRC 2012 validation set is used to evaluate the impact of batch size. A synthetic dataset with varying resolutions was utilized to evaluate the impact of image size. Images were resized to the desired resolution and subsequently saved in JPEG format. 

Throughout the experiment, a consistent preprocessing routine was applied to each image. Images will first be resized to $256$x$256$ pixels, then central crop to $224$x$224$ pixels. Then, images were normalized using a mean of $[0.485, 0.456, 0.406]$ and a standard deviation of $[0.229, 0.224, 0.225]$. Finally, images were cast to the Float32 data type.

\textbf{Model:}
During the experiments, three different sizes of models were used to represent models with varying levels of computational demands. These models were ResNet18, ResNet50, and ResNet152. ResNet is a widely-used convolutional network structure known not only for object classification tasks, but also as the backbone for other models like object detection models. 

All models and their weights were obtained from Torchvision~\cite{torchvision} and exported to the platform-neutral ONNX format using PyTorch. The TensorRT command-line wrapper tool trtexec~\cite{tensorRT} was used to build the TensorRT engine. Due to distinct hardware architecture, GPU and DLA engines were built separately. For efficient inference and alignment with the practical usage scenarios on edge devices, INT8 quantization was performed.

\subsection{Comparing Preprocessing Pipelines}
\label{sec:ComparingPreprocessingPipelines}

\begin{figure}[!t]
     \centering

    \centering
    \includegraphics[width=\columnwidth]{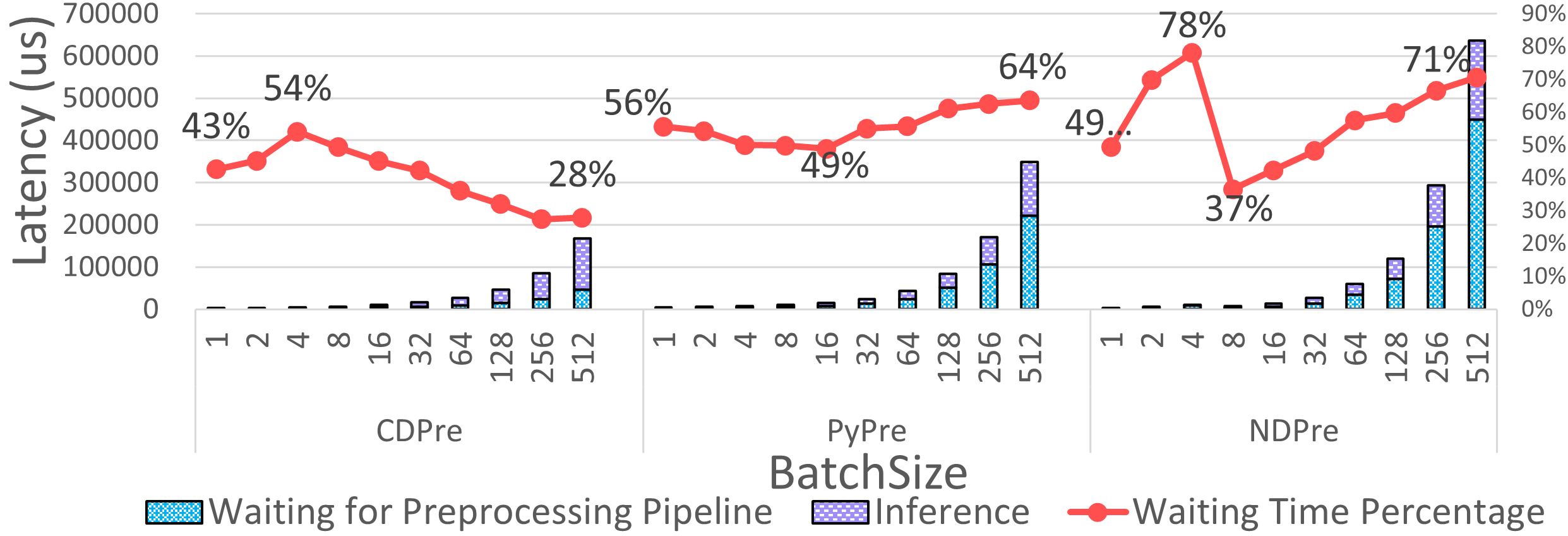}
    \caption{
    Inference batch latency of CDPre, PyPre, and NDPre pipelines with varying batch sizes. With percentage of time inference pipeline waiting for the preprocessing pipeline.}
    \label{fig:BSChart_lat}

\vspace{-1ex}
\label{fig:BSChart}
\end{figure}

In this experiment, the latency was compared when changing the pipeline among \pypre{}, \cdpre{}, and \ndpre{}. Additionally, the batch size was varied to determine its impact on the pipeline.

As depicted by Figure~\ref{fig:BSChart_lat}. \cdpre{} consistently delivers better throughput than \pypre{} and \ndpre{}. The primary advantage of \cdpre{} arises from utilizing the GPU during the image transformation phase. 
However, launching GPU kernels introduces overhead, but as batch size increases — yet before running out of memory — this overhead becomes amortized, leading to improved performance. For \ndpre{}, the parallelism of NVJPEG is not as extensive as that of the CPU. Besides, The fine-grain pipeline design with intermediate buffer design makes \ndpre{} less sensitive to BS changes.

This experiment demonstrates the benefits of offloading image transformation to the GPU. \ndpre{}'s parallelism and intermediate buffer design make it less sensitive to changes in BS, but it performs poorly at excessively large or small BS. Additionally, NVJPEG does not offer improved performance for datasets with significant variations in image size and ratio.

\vspace{-2ex}

\subsection{Preprocessing Time Breakdown Evaluation}
\label{sec:breakdownEval}

\begin{figure}[!t]
     \centering

\begin{subfigure}[b]{\columnwidth}
    \centering
    \includegraphics[width=\linewidth]{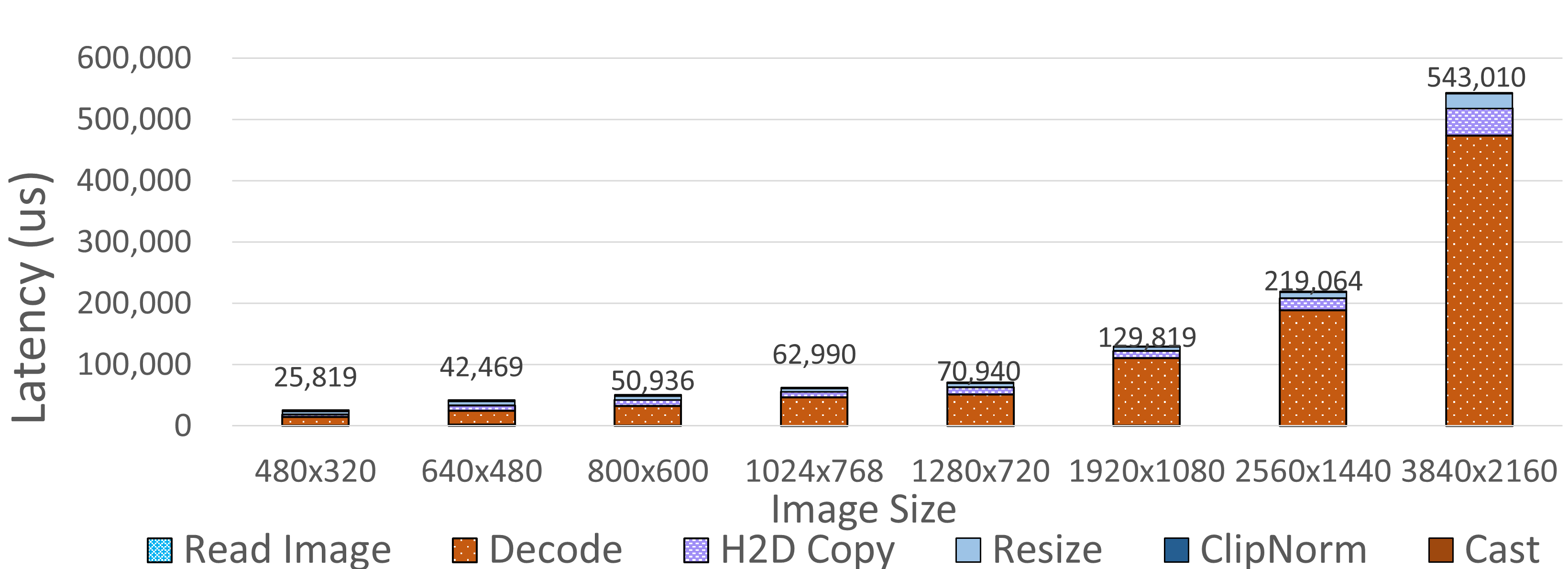}
    \vspace{-3.5ex}
    \caption{CD-Pre time breakdown}
    \label{fig:Time_Composite_CD}
\end{subfigure}

\begin{subfigure}[b]{\columnwidth}
    \centering
    \includegraphics[width=\linewidth]{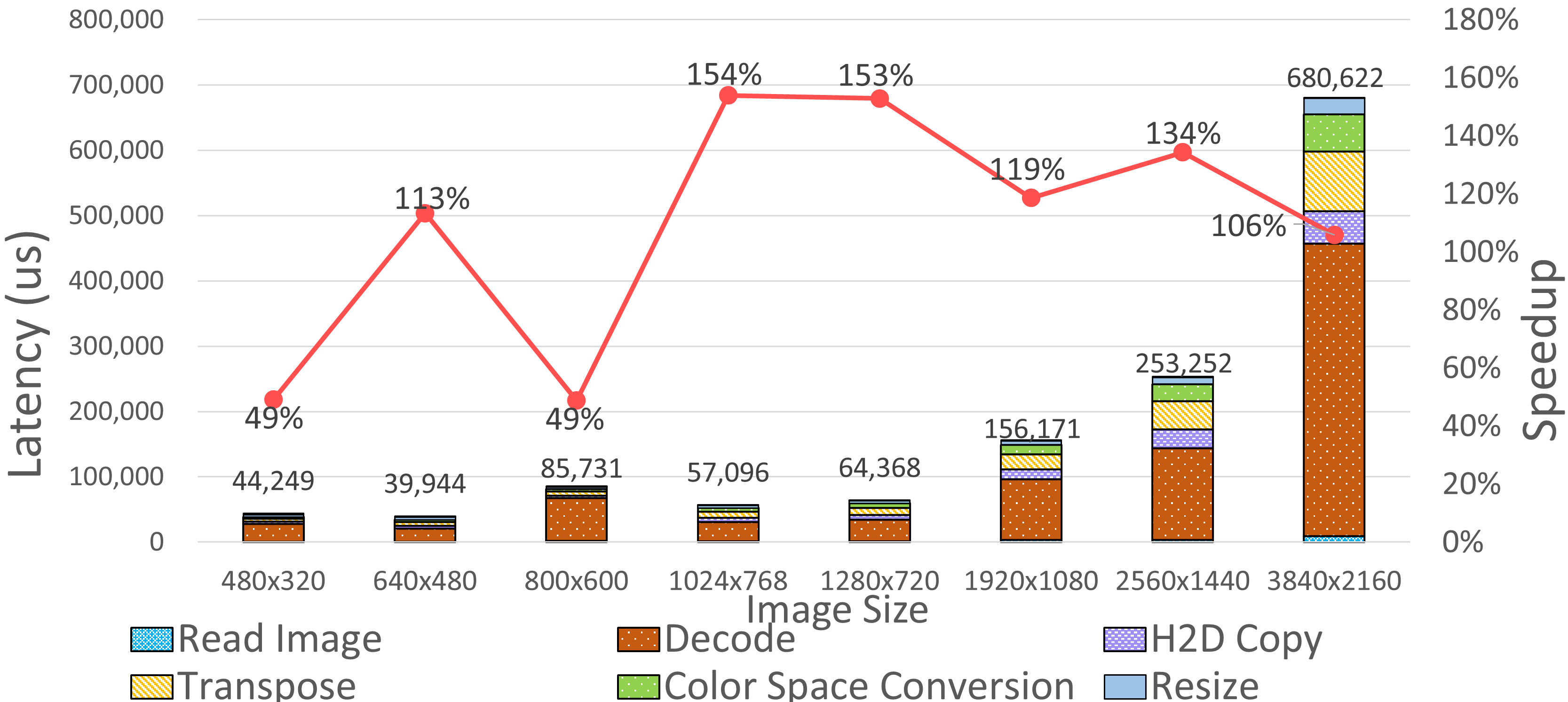}
    \caption{ND-Pre time breakdown. With decoding speedup compared with \cdpre{} under the same image size.}
    \label{fig:Time_Composite_ND}
\end{subfigure}

\caption{Time breakdown of Preprocessing Pipelines. ResNet50 with a batch size of 64 and DLA Inference.}

\label{fig:Time_Composite}
\vspace{-2ex}
\end{figure}

To analyze the time composition of each preprocessing pipeline. Following the formula in Section~\ref{sec:design:timeline:formula}. The inference was offloaded using the DLA unit to minimize the impact of GPU activity on the preprocessing pipeline. The results are presented in Figure~\ref{fig:Time_Composite}.

In both subfigures of \cdpre{} and \ndpre{}, it's evident that decoding time constitutes a major portion of the preprocessing phase. In Fig.~\ref{fig:Time_Composite_CD}, for an image size of 480x320, decoding accounts for \percent{53} of the total time. This percentage increases as the image size grows, reaching as much as \percent{87} for 3840x2160 image size.

In the Fig.~\ref{fig:Time_Composite_ND}, the proportion of time taken for decoding remains relatively stable as the resolution increases. However, color space conversion and accompanying transpose also occupy a non-negligible amount of time, accounting for \percent{10} and \percent{17} respectively at image size of 2560x1440. Comparing the decoding time of \ndpre{} and \cdpre{} at the same image sizes, as shown in the figure as Decoding Speedup, it is observed that \ndpre{} has up to a \percent{54} advantage over CPU decoding for image sizes larger than 1024x768. Still, it is less efficient for smaller image sizes. 
Though \ndpre{} owns an advantage in decoding, when adding up additional GPU color space conversion overhead, \ndpre{} does not show a latency advantage over \cdpre{} in single instance scenarios.

\vspace{-0.5ex}

\subsection{Inference Devices Comparison}
\label{sec:eval:devicesComparison}

\begin{figure*}[t!]
    \centering
    \vspace{-1ex}
    \includegraphics[width=1.0\textwidth]{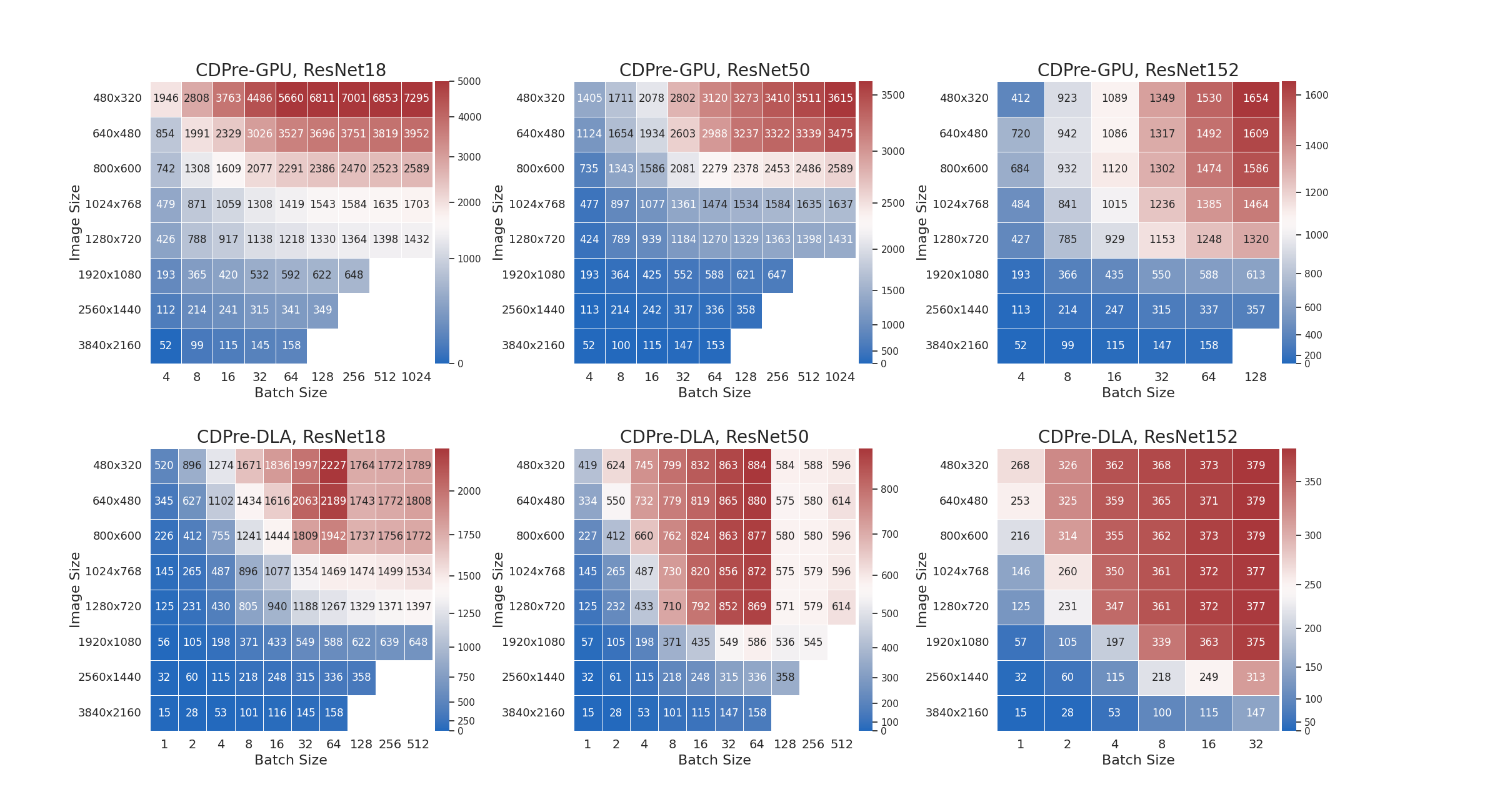}
    \caption{Inference throughput for GPU and DLA over different Models, BatchSizes and Image Sizes.
The first row uses the GPU for inference, the second row uses the DLA. In each subplot, color indicates the throughput for the combination.}
    
    \vspace{-2ex}
    \label{fig:heapmap}
\end{figure*}

In this section, a comparison of throughput between the \cdpre{} and \cdpredla{} pipelines was conducted to evaluate the differing characteristics of the GPU and DLA during inference.

Figure~\ref{fig:heapmap} reveals that for GPU-based inference, when dealing with image sizes smaller than 1280x720, scaling up the batch size leads to a corresponding increase in throughput until it gradually saturates. This is consistent with the previous conclusion that the \cdpre{} preprocessing pipeline prefers larger batch sizes. For image sizes larger than 1280x720, using a large batch size can lead to out-of-memory (OOM). Additionally, the throughput will be constrained by a preprocessing bound. Consequently, for three models with different inference latencies, a similar throughput is observed.

In the second row, the DLA-based inference, increasing the batch size does not necessarily enhance overall throughput. The DLA, being a less powerful computing unit compared to the GPU,  with its own microcontroller and hardware buffers, exhibits different performance characteristics. The results indicate that for ResNet18 and ResNet50, the throughput peaks at a batch size of 64, after which there is a drop in performance. When the image size is increased, the same preprocessing bound observed in the corresponding GPU cases evident. For example for 2560x1440 image size with ResNet50, \cdpre{} and \cdpredla{} has very close number of $358$ images/sec. So that in preprocessing bounded case, DLA and GPU based inference will lead to similar throughput.

\subsection{Multi-Instance Design with NVJPEG Involved}

\subsubsection{Characteristic of Multi-Instance}

\begin{table}[ht]
\centering
\resizebox{0.90\columnwidth}{!}{%
\begin{tabular}{|c|c|c|c|c|}
\hline
\textbf{\#Instances} & \textbf{CDPre} & \textbf{CDPre-DLA} & \textbf{NDPre} & \textbf{Images/sec} \\

\hline

\multirow{3}{*}{1}  
&  1  &	     &      &	\makecell{\textcolor{red}{\textbf{358.02 Single-instance Baseline}} }\\\cline{2-5}
&      &	 1  &      &	 336.44 \\\cline{2-5}
&      &	     &  1  &	 375.07 \\\cline{2-5}
\hline
\hline

\multirow{4}{*}{2}
&  2  &	     &      &	 382.98 \\\cline{2-5}
&  1  &	     &      &	 370.61 \\\cline{2-5}
&      &	     &  1  &	 477.10 \\\cline{2-5}
&  1  &	     &  1  &	 490.20 \\\cline{2-5}
\hline
\hline

\multirow{4}{*}{3}
&  1  &	 2  &      &	 378.22 \\\cline{2-5}
&  3  &	     &      &	 379.33 \\\cline{2-5}
&      &	 2  &  1  &	 489.88 \\\cline{2-5}
&  2  &	     &  1  &	 \textcolor{red}{\textbf{498.14 Best}} \\\cline{2-5}

\hline
\hline

\multirow{5}{*}{4}
&  4  &	     &      &	 382.29 \\\cline{2-5}
&      &	 4  &      &	 366.14 \\\cline{2-5}
&      &	 3  &  1  &	 453.92 \\\cline{2-5}
&  3  &	     &  1  &	 442.91 \\\cline{2-5}
&  1  &	 2  &  1  &	 472.09 \\\cline{2-5}

\hline
\hline

\multirow{3}{*}{5}
&  5  &	     &      &	 374.67 \\\cline{2-5}
&  1  &	 4  &      &	\makecell{\textcolor{red}{\textbf{383.10 Best without NVJPEG}} }\\\cline{2-5}
&      &	 4  &  1  &	 443.02 \\\cline{2-5}
\hline
\hline

\multirow{3}{*}{6}
&  6  &	     &      &	382.34  \\\cline{2-5}
&  2  &	 4  &      &	 380.76 \\\cline{2-5}
&  3  &	 2  &  1  &	 445.63 \\\cline{2-5}
\hline
\hline

\multirow{4}{*}{7}
&  1  &	 6  &      &	 347.16 \\\cline{2-5}
&  3  &	 4  &      &	 372.62 \\\cline{2-5}
&      &	 6  &  1  &	 366.62 \\\cline{2-5}
&  2  &	 4  &  1  &	 437.47 \\\cline{2-5}
\hline

\end{tabular}
}
\caption{
Throughput comparison for various Multi-Instance combinations, ResNet50 with 2560x1440 images used. Column $1$ indicates the total number of instances involved. Column $2\sim4$ number indicates the number of that kind of instance involved.
}
\vspace{-4ex}
\label{tab:multi-instance}
\end{table}

In this section, the effectiveness of the Multi-Instance design was evaluated, incorporating various combinations of pipelines with and without the involvement of the NVJPEG hardware unit. The experiment was conducted using the ResNet50 model with an image size of 2560x1440, which is a preprocessing bounded scenario.

By incrementally increasing the total number of instances, an exhaustive test of possible combinations was carried out. Notice that the Orin device is equipped with two DLA units. Therefore, instances of the \cdpredla{} pipeline will be evenly distributed across these two DLAs. Additionally, based on the result of Experiment~\ref{sec:eval:devicesComparison}, the batch size for the DLA based processing is set to 64 to maximize performance. For the GPU, the batch size is set as large as possible until it runs out of memory. 

For Table~\ref{tab:multi-instance}. Firstly, compare combinations where only one type of pipeline is involved, but varying the number of instances. With a single \cdpre{} instance, the throughput was $358.02$ images/sec. With two instances, this increased to $382.98$ images/sec, indicating improved utilization. However, adding more instances did not lead to further performance gains. Even adding to six instances merely have $382.34$ images/sec.

In scenarios involving DLA offload for inference, similar throughput was observed comparing its fully GPU based counterpart. For example, for Multi-Instance design consist with three instances, one \cdpre{} and two \cdpredla{} instances combination delivers $378.22$ images/sec throughput. While for combination with three \cdpre{} instances, throughput is $379.33$ images/sec. In preprocessing bounded situations, these combinations offered comparable performance, and additional inference devices did not enhance overall throughput.

However, when considering the inclusion of an \ndpre{} instance, the dynamics changed. Although a single \ndpre{} instance did not show a significant advantage over a single \cdpre{} instance ($375.7$ images/sec vs. $358.02$ images/sec), combinations involving \ndpre{} demonstrated higher performance in Multi-Instance settings.  Under two \cdpre{} and one \ndpre{} combination, achieved peak throughput of $498.14$ images/sec, marking an approximate \percent{30.02} improvement over the highest Multi-Instance throughput without \ndpre{} ($383.10$ images/sec).

\begin{figure}[htbp!]
     \centering
\begin{subfigure}[b]{1.0\columnwidth}
    \centering
    \includegraphics[width=\linewidth]{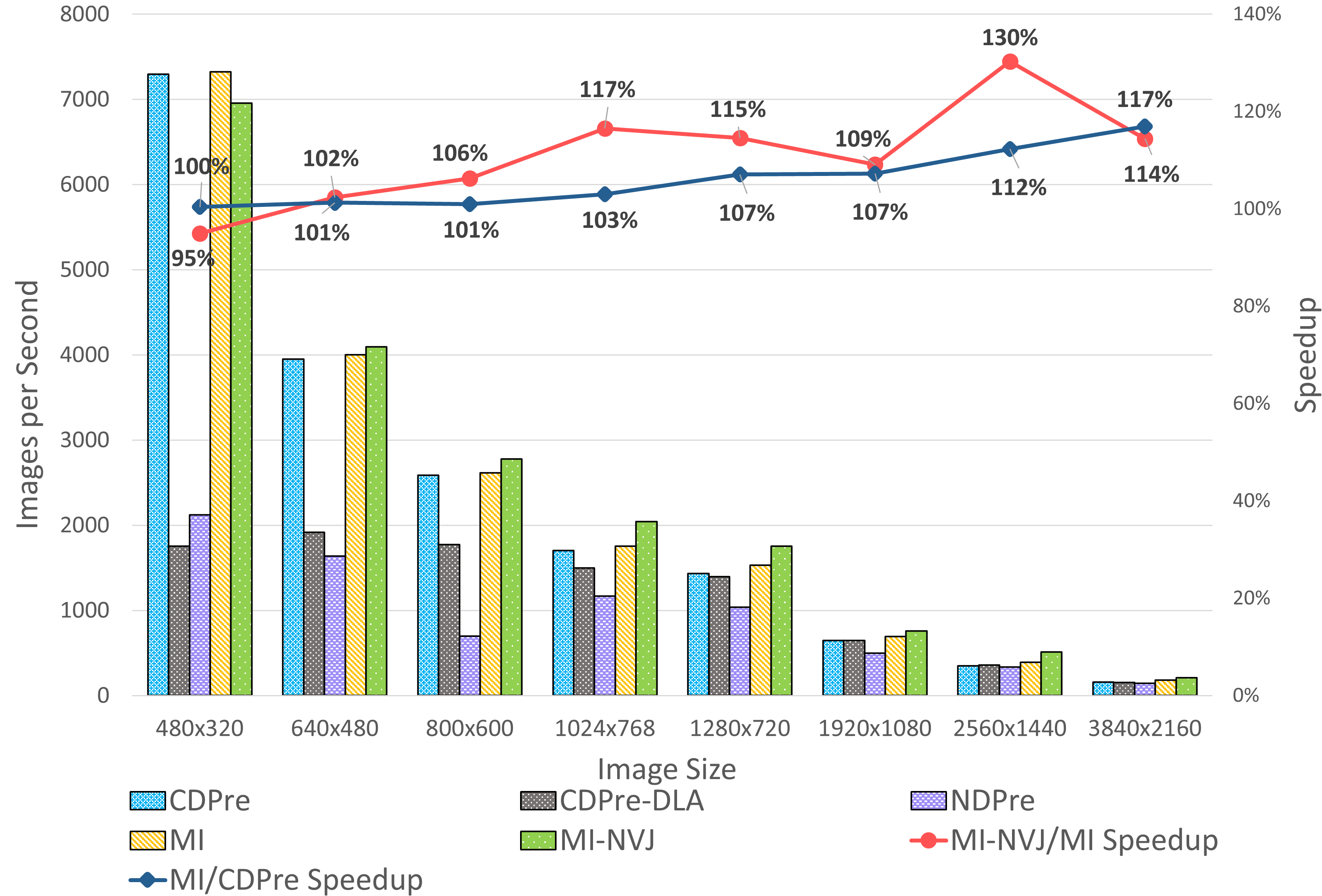}
    \caption{ResNet18}
    \label{fig:main_result_18}
\end{subfigure}

\begin{subfigure}[b]{1.0\columnwidth}
    \centering
    \includegraphics[width=\linewidth]{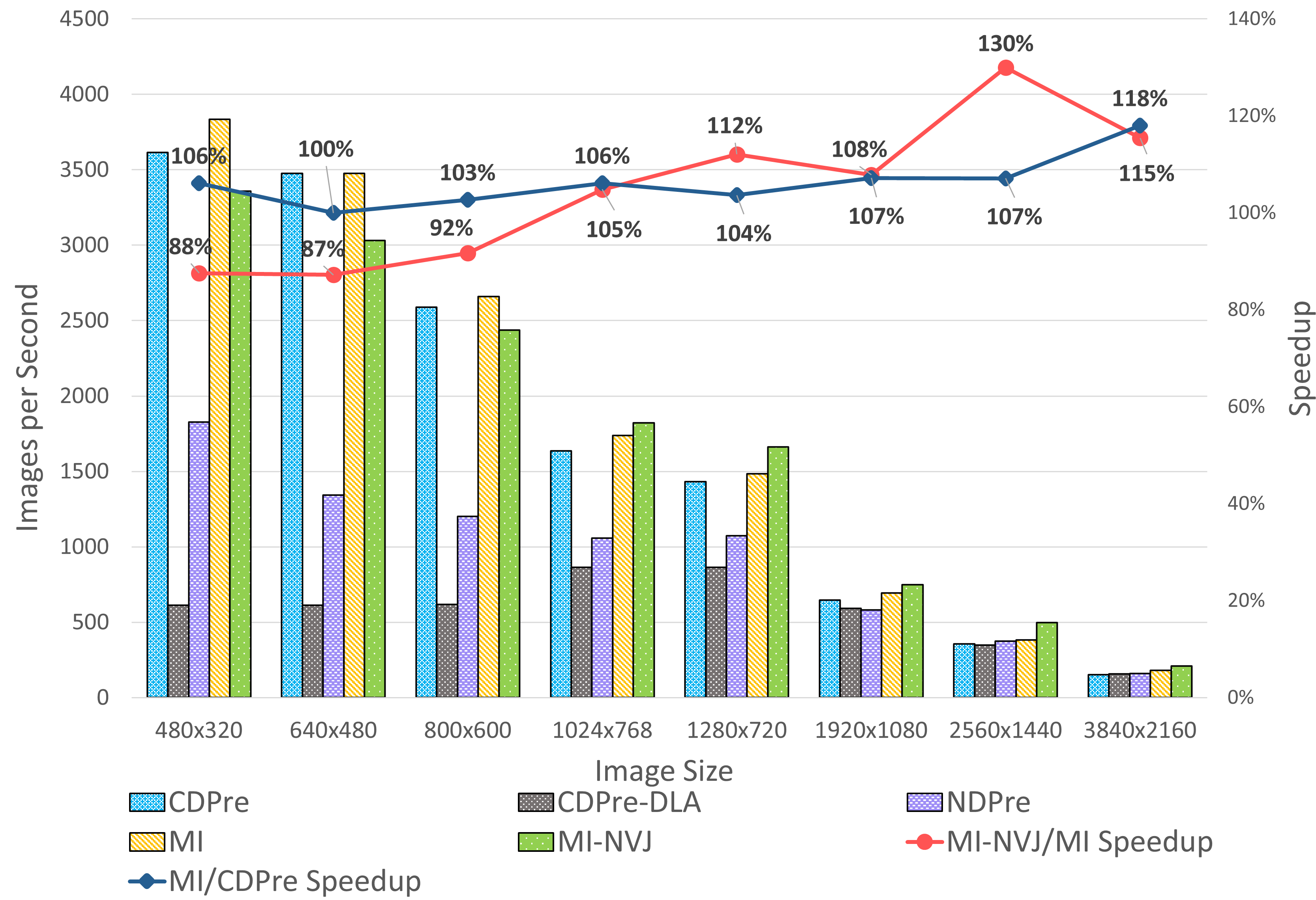}
    \caption{ResNet50}
    \label{fig:main_result_50}
\end{subfigure}

\begin{subfigure}[b]{1.0\columnwidth}
    \centering
    \includegraphics[width=\linewidth]{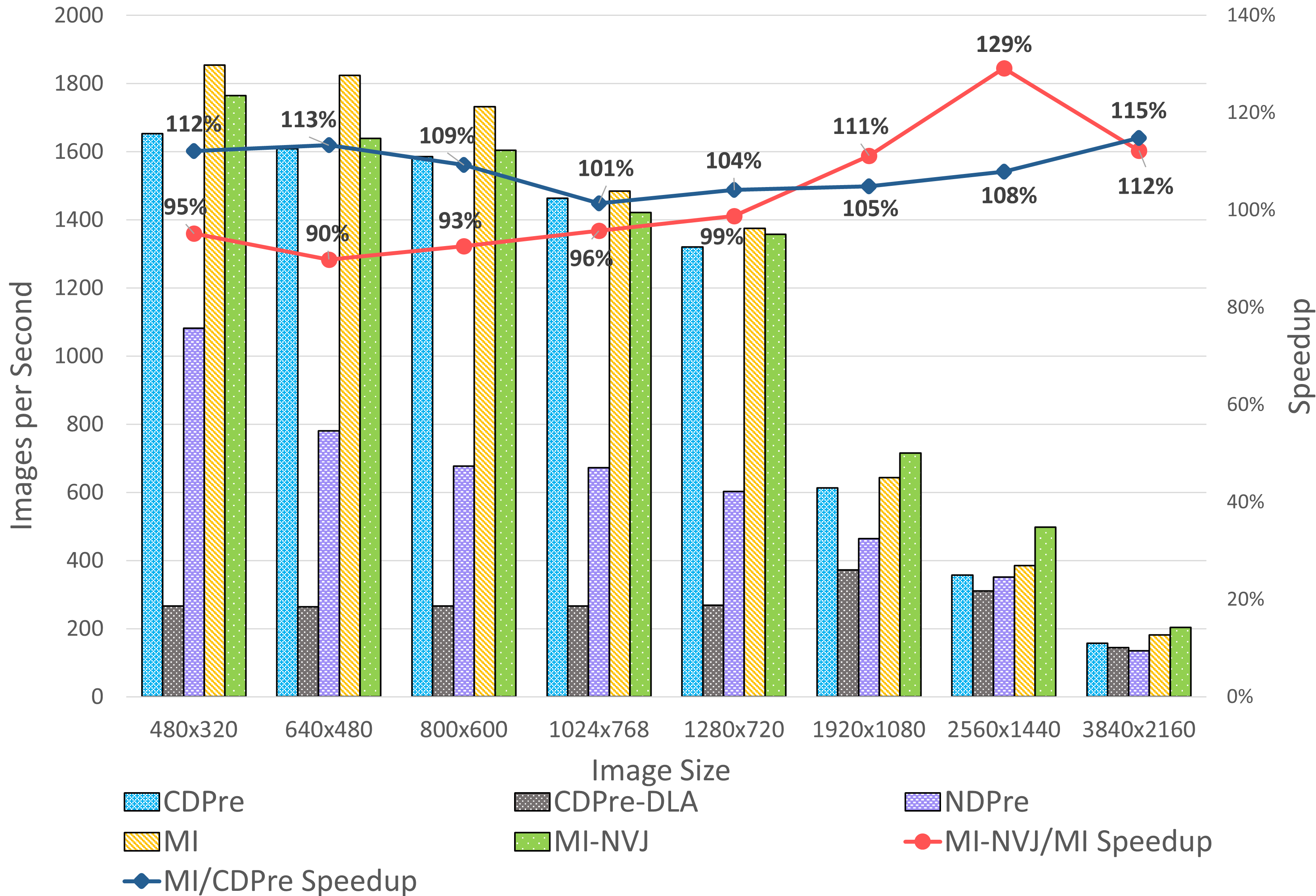}
    \caption{ResNet152}
    \label{fig:main_result_152}
\end{subfigure}

\vspace{-1.2ex}
\caption{Inference throughput of various pipeline designs. All configurations tuned with optimal batch size. MI and MI-NVJ optimized with best instance combinations. `MI-NVJ/MI Speedup' demonstrates the acceleration achieved by integrating NVJPEG. `MI/CDPre Speedup' highlights the advantages of employing a Multi-Instance design.
}
\vspace{-1.5ex}
\label{fig:main_result}
\end{figure}

\subsubsection{Multi-Instance Tunning}

Given that a \minv{} leads to improved performance in preprocessing bounded situations, this section extends the comparison to ResNet18, ResNet50, and ResNet152 models across varying image sizes.

From Figure~\ref{fig:main_result_18}, it shows that utilizing the \mi{} design can offer a speedup of up to \percent{18.0} for ResNet18 at 3840x2160 resolution when compared to  \cdpre{}. The \mi{} design is particularly beneficial for large image sizes, which typically represent preprocessing-bounded scenarios.

Additionally, the \mi{} design demonstrates potential benefits in inference-bounded situations, since it can paralleling two DLAs and a GPU for inference. For example in Fig.~\ref{fig:main_result_152}, with ResNet152 at 640x480 resolution, there is a \percent{13.3} advantage over  \cdpre{}.

\minv{}, compared to \mi{}, mandates the involvement of NVJPEG to offload JPEG decoding, which generally delivers better performance than MI for large image sizes. For example in Fig.~\ref{fig:main_result_50}, in the case of ResNet50 at 2560x1440, \minv{} outperforms \mi{} by \percent{30.02}. However, for smaller image sizes, this design may lead to a reduction in overall throughput.

Compared to  \cdpre{}, for various models and image sizes, both \mi{} and \minv{} designs enhanced device utilization, or offloaded decoding/inferencing to more hardware devices, leading to improved throughput. However, the involvement of the NVJPEG hardware unit can cause a degradation in performance with smaller image sizes. Similarly, DLA does not always contribute to an increase in overall throughput.

\section{Related Work}
\label{sec:related_work}

There are several related works concerning edge device inference, GPU-accelerated JPEG decoding, and optimizing preprocessing pipelines. Jeong et al.\cite{jeong_tensorrt-based_2022} proposed a TensorRT-based parallel inference framework for Jetson devices. Performing model/data parallelism on GPU and DLAs. 
Cheng et al.\cite{cheng_dlbooster_2019} proposed a data preprocessing pipeline that offloads decoding to either a GPU or FPGA. 
Weissenberger and Schmidt~\cite{weisenberger_accelerating_2021} introduced methods for accelerating JPEG decompression on GPUs. Wang et al.~\cite{wang_accelerating_2020} demonstrated the benefits of using GPU-assisted image decoding in computer vision deep learning workflows. These works highlight the importance of JPEG decoding in deep learning workflows and the advantages of offloading these tasks to GPUs. 
However, their focus was on preprocessing in server platforms with powerful GPUs and did not consider preprocessing and inference as an integrated whole, as we have done with edge devices. To the best of our knowledge, our work is the first to utilize \nvjL on the Jetson Platform for offloading decoding. At the same time, optimized inference by utilizing all hardware units, including the CPU, GPU, DLA, and NVJPEG, in a parallel manner.

\section{Conclusions}
\label{sec:conclusion}

This work introduces the use of \nvjL{} combined with a fine-grained pipeline design for offloading JPEG image decoding in edge inference workflows. Additionally, the implementation of the \minvL{} design enables full overlap of NVJPEG and CPU decoding while utilizing both GPUs and DLAs for inference. Our comprehensive evaluation highlights the distinct characteristics between CPU and NVJPEG, as well as GPU and DLA. The assessment of the \minv{} design reveals that incorporating \nvjL{} can lead to a performance advantage of up to \percent{30.02}  compared to the highest throughput achievable by \mi{} without NVJPEG.  

Based on our findings, we propose that users, prior to deploying a specific model with a given data source, first determine the optimal batch size for DLA. Subsequently, they should adjust the GPU batch size to its maximum without run-out-of-memory. Then, by conducting a profiling run over a few batches, they can explore multi-instance designs involving various hardware units, ultimately selecting the most optimized combination for their workflow.

\hypersetup{linkcolor=black}

\bibliographystyle{ieeetr}
\bibliography{bibs/main,bibs/nvjWork}

\end{document}